\documentclass[preprint2,tighten,times]{aastex6}
\pdfoutput=1 
\usepackage{hyperref,verbatim}
\usepackage{amsmath,amstext,mathrsfs}
\usepackage[all]{hypcap} 
\usepackage{multirow}
\usepackage{xcolor}
\usepackage{marginnote}
\usepackage{subfigure}

\begin{document}
	\title{Intensity Gradients Technique: Synergy with Velocity Gradients and Polarization Studies} 
	\author{Yue Hu\altaffilmark{1,2}, Ka Ho Yuen\altaffilmark{2}, A. Lazarian\altaffilmark{2}}
	\email{yue.hu@wisc.edu;\\
		kyuen2@wisc.edu;\\
		alazarian@facstaff.wisc.edu}
	\altaffiltext{1}{Department of Physics, University of Wisconsin-Madison}
	\altaffiltext{2}{Department of Astronomy, University of Wisconsin-Madison}

\begin{abstract}
Magnetic fields are ubiquitous in the interstellar medium but are notoriously difficult to study through observation. Making use of the advances in our understanding of MHD turbulence and turbulent reconnection, the Velocity Gradient Technique (VGT) was suggested and successfully applied to study magnetic fields utilizing spectroscopic data. Applying the tools developed for VGT to intensity statistics, we introduce the Intensity Gradients Technique (IGT) as a complementary tool that can be used synergistically with VGT. In this paper, we apply IGT to a diffuse \ion{H}{1} region selected from the GALFA-\ion{H}{1} survey and compare the intensity gradient maps with those obtained using velocity gradients as well as Planck polarization measurements. We demonstrate the possibility of using IGT and VGT for both studying the magnetic field and identifying shocks in the diffuse interstellar medium. We also explore the ability of IGT in locating self-gravitating regions and calculating Alfvenic Mach number, both alone and in combination with VGT and polarimetry. We compare IGT with the Histogram of Relative Orientation (HRO), which utilizes intensity gradients to characterize the relative orientation of column density structures and local magnetic fields.
\end{abstract}
	
\section{Introduction}
The magnetic force is second in importance, after gravity, in the present-day understanding of the Universe \citep{Sp78, Shu83, M91,K12}. In an astrophysical setting, magnetic fields are embedded in turbulent conducting plasmas \citep{1981MNRAS.194..809L,2004ApJ...615L..45H,2010ApJ...714.1398C,2004ARA&A..42..211E,2007ARA&A..45..565M}, making MHD turbulence an accurate description of the state of astrophysical fluids \citep{BL19}. To understand the critical astrophysical process, e.g., the process of star formation \citep{1956MNRAS.116..503M,2006ApJ...646.1043M,2006ApJ...647..374G,2007ApJ...664..975J}, it is essential to know both the properties of the turbulent magnetic field and the density of the matter. In particular, it is crucial to know the density enhancement arising from shocks and self-gravitation.

Studies of the magnetic field in cold diffuse gas and molecular clouds commonly employ the starlight polarization and thermal emissions produced by aligned grains \citep{Andersson2015InterstellarAlignment} as well as molecular line splitting (Zeeman effect) from radio to optical wavelengths
\citep{2010ApJ...725..466C,2012ARAA...50..29}. Far-infrared dust polarization measurements can not only determine the direction of the projected magnetic field $B_{POS}$ but also roughly estimate the magnetic field strength through the Davis-Chandrasekhar-Fermi method \citep{1953ApJ...118..113C,1951PhRv...81..890D}. 

However, measurements of the magnetic field using far-infrared dust polarization utilize ground-based telescopes, which are affected by the radiative absorption that happens as radiation passes through the Earth's atmosphere. Moreover, the reliability of magnetic field tracing obtained using polarization techniques decreases when grain alignment and Radiative Torques are weakened by light extinction \citep{2007MNRAS.378..910L}, e.g., in molecular clouds at high optical depths \citep{Andersson2015InterstellarAlignment}. Although line splitting, such as the Zeeman effect, can directly measure the strength of the line-of-sight magnetic field $B_{LOS}$ \citep{2010ApJ...725..466C}, high sensitivity requirements and long integration times limit the applicability of Zeeman measurements.

The Velocity Gradients Technique (VGT, \citealt{GL17, YL17a, YL17b,2018ApJ...853...96L, 2018ApJ...865...46L}) was developed as a new method to trace the direction of magnetic fields by using spectroscopic data. The theoretical basis of the technique, discussed in Sec.~\ref{sec:theory}, is the theory of MHD turbulence and turbulent reconnection. The utility of VGT technique has been successfully tested through numerical simulations and through comparison with magnetic field morphology predictions of diffuse ISM and molecular clouds obtained using polarimetry  \citep{2018ApJ...853...96L,2018ApJ...865...46L,2018MNRAS.480.1333H,PGCA,2017arXiv170303035G,2018arXiv180909806H,survey, Laura}. In addition to tracing magnetic fields, VGT is also a powerful technique of obtaining the media magnetization level \citet{2018ApJ...865...46L}, sonic Mach number measurements \citep{2018arXiv180200024Y}, and provides a statistical error measure.

The intensity of emissions from both gas and dust provides additional information about the ISM which is different from the information provided by VGT. It is therefore attractive to investigate the new insight provided by the intensity gradients. Our goal is to explore the information obtainable through the synergy of VGT and the Intensity Gradients Technique (IGT) and IGT on its own.

\cite{Soler2013} proposed the technique termed Histogram of Relative Orientations (HRO), to characterize the relative orientation of density gradients and local magnetic fields. However, HRO is implemented in a way and with the goals that are radically different from VGT. HRO is not a technique for tracing magnetic fields, but rather one for exploring the statistics of the change of relative orientations of intensity gradients and magnetic fields in response to changes in column densities. The authors rely on polarization measurements to find magnetic field orientation. VGT, in contrast, explores the pointwise statistics of the magnetic field and does not require any outside measurements. In addition, some of VGT ideas and approaches were successfully borrowed and implemented within HRO as it was evolving (see \citealt{Soler2018}).

In what follows, we illustrate the theoretical foundation of IGT in Sec.~\ref{sec:theory}. In Sec.~\ref{Sec.MHD data}, we describe the MHD simulation data used in this work. In Sec.~\ref{sec.method},  we briefly describe the algorithms used in the implementation of IGT. In Sec.~\ref{sec:simulation} and Sec.~\ref{sec.obs}, we show our results obtained using IGT in numerical simulations and observations, respectively. In Sec.~\ref{sec.dis}, we discuss the possible application of IGs with the latest development of VGT. In Sec.~\ref{sec.con}, we give our conclusions.

\section{Theoretical Motivation and Expectation}
\label{sec:theory}
\subsection{Theory of MHD turbulence and VGT}
	
The theoretical justification why velocity gradients trace magnetic field comes from the theories of MHD turbulence as well as turbulent reconnection. The theory of MHD turbulence has been given a boost by the prophetic study by \citet{GS95}, denoted as GS95 later. In particular, \cite{GS95} predicted the turbulent eddies to be anisotropic and showed that the degree of turbulence anisotropy increases as the scale of turbulent motions decreases. The subsequent study of turbulent reconnection in \citet{LV99} demonstrated that turbulent reconnection of magnetic field is an intrinsic part of the MHD turbulent cascade. The eddies perpendicular to magnetic field direction evolve freely with magnetic reconnection taking place over just one eddy period. As a result, anisotropic eddies are aligned with the direction of the magnetic field in their direct vicinity, i.e. the {\it local} magnetic field direction. The latter is absolutely crucial element for VGT technique, as it testifies that the small velocity fluctuations are well aligned with the local direction of magnetic field.\footnote{The derivations in \cite{GS95} for the anisotropy are done using the mean field reference frame. In fact, the  GS95 scaling are not valid in this frame of reference.} This phenomena has been confirmed by the numerical studies in \citet{2000ApJ...539..273C} and \citet{2001ApJ...554.1175M}.

Employing the notion of fast turbulent reconnection , it is obvious that the motions of eddies with size $l_{\bot}$ perpendicular to the local direction of magnetic field are not constrained by magnetic field. Thus, they should exibit hydrodynamic-type statistics, i.e. obey the Kolmogorov law $v_{l,\bot}\sim l_\perp^{\frac{1}{3}}$, where $v_{l,\bot}$ is the turbulence's injection velocity perpendicular to  the local direction of magnetic field. By equating the period of Alfv\'{e}nic wave  and turbulent eddy's turnover time :
\begin{equation}
    \frac{l_{\|}}{v_A}=\frac{l_{\bot}}{v_{l,\bot}}
\end{equation}
where $v_A$ is the Alfvenic velocity. One can obtain the relation between the long and short axes of the eddies \citep{LV99}:
\begin{equation}
    l_\|\sim l_{\perp}^{\frac{2}{3}}
\end{equation}
In this paper, we will refer to the above expression for $v_{l,\bot}$ and the relation between $l_\|$ and $l_\bot$ as GS95 relations. Note, that the anisotropy relation is not valid in the reference system of the mean magnetic field. The latter, in fact, was a frequent mistake of many researchers who tried to measure the scale dependent anisotropy both from numerical simulations and observations.
	
In terms of VGT, the Kolmogorov scaling means that (1) the gradients of velocity amplitude scale as $v_{l,\bot}/l_{\bot}\sim l_{\bot}^{-2/3}$, i.e. the smallest resolved scales are most important in calculating the gradients; (2) the measured velocity gradients are perpendicular to the magnetic field at the smallest resolved scales, i.e. they well trace the magnetic field in the turbulent volume. Similar to the case of far-infrared polarimetry, one should turn the direction of gradients by  90$^\circ$ to obtain the magnetic field direction.
	
\subsection{MHD turbulence and density statistics}
	
In MHD turbulence, velocity and magnetic field fluctuations follow the same GS95 relations for Alfvenic part, which is a dominant part of the MHD cascade \citep{Cho2002, Cho2003CompressibleImplications,2001AAS...198.9003L}. The situation is more complicated for the density field. In fact, in \citet{2005ApJ...624L..93B}, it was shown that for supersonic turbulence, the GS95 relations could be valid for low-value density enhancements, while the relation becomes different for high-value density fluctuations. Further studies, e.g., \citet{2007ApJ...658..423K, YL17b,2019arXiv190506341X} show that the high contrast density fluctuations are created by shocks perpendicular to the local direction of the magnetic field. These structures do not obey the GS95 relations. Therefore the study of density gradients can provide additional information that is not reflected by velocity gradients.

In particular, \cite{2007ApJ...658..423K} numerically studied subsonic turbulence with the presence of a relatively strong magnetic field. They showed that the spectrum of density scales is similar to the pressure, i.e., $E\sim k^{-7/3}$. This scaling type is theoretically expected for the polytropic equation of state $p\propto \rho^{\gamma}$, where p is the pressure, $\rho$ is density, and $\gamma$ is polytropic coefficient \citep{2003matu.book.....B}. As for super-sonic turbulence, the density spectrum becomes shallower because shocks accumulate the fluid into the local and highly dense structures. However, \citet{2005ApJ...624L..93B} shows that by filtering out high contrast density clumps, the density statistics still exhibit Kolmogorov-type scaling $E\sim k^{-5/3}$ and scale-dependent anisotropy of the GS95 type $l_{||}\sim l_{\perp}^{\frac{2}{3}}$.
	
\subsection{Observations of velocity and density fluctuation}
	
Velocities statistics are not directly available from observations. To get insight into velocity statistics, the traditional way is to use velocity centroids \citep{2005ApJ...631..320E}. Those were, in fact, first used for the velocity gradient studies \citep{GL17,YL17a,YL17b}. Later \citet{LP00,LP04} developed the theory of statistics of the Position-Position-Velocity (PPV) spectroscopic data cubes and \citet{2017MNRAS.464.3617K} elaborated the theory. Based on these theories, it was proposed to use fluctuations of intensity within thin velocity channel maps to trace the velocity gradients \citep{2018ApJ...853...96L}.
	
Similarly, the observations, as a rule, do not provide 3D density distributions, but column densities. For instance, due to the high degree of mixing of dust and gas, the far-infrared emission of dust reflects the column densities of diffuse interstellar gas. However, the column density information can also be obtained from integrating the spectroscopic data over the line-of-sight velocities. This way of studying is advantageous as it allows us to separate the contributions of different volumes of emitting/absorbing gas along the line of sight. Therefore in what follows we focus on obtaining the column density information from the spectroscopic data. 
	
\subsection{Density fluctuations in thick channel maps}
	
Three dimensional MHD turbulence data, i.e., in Position-Position-Position (PPP) space, is not available in observation, but \citet{LP00} explored the possibility of using the statistics of velocity fluctuations in PPV cubes to study turbulence. The subsequent works \citep{KLP16,KLP17b} used PPV cubes to detect the anisotropy of velocity distribution that is induced by magnetic field. However, the information conversed from PPP to PPV  we see in observation is not trivial, especially about how the density and velocity structures are modified. 
	
\citet{LP00} first proposed the concept of velocity caustics to signify the effect of density structure distortion due to turbulent velocities along the line of sight. Since the density structure with different velocities is sampled into different velocity channels, the density structure is significantly modified. In \citet{LP00}, the significance of velocity caustics in PPV cubes is quantified in terms of the density spectral index, which the latter highly dependent on the sonic Mach number M$_s$ \citep{Cho2002,Cho2003CompressibleImplications}. When the density power spectrum is steep, i.e., $k<-3$, the emissivity spectrum of PPV cube is dominated by the velocity fluctuation. Thus for such flows, the density fluctuations in thin channels of PPV data are following the turbulent velocity statistics, while the dominance of velocity fluctuation will lead to a shallower emission spectrum if the velocity channels are sufficiently thin. Later studies \citep{LP04, LP06,LP08} revealed that the same classification is also seen in absorption media and this has been extensively applied to observations \citep{1993MNRAS.262..327G, LP06,2000ApJ...543..227D,LP04,2001ApJ...551L..53S,2001ApJ...561..264D,2006ApJS..165..512K,2006AIPC..874..301L,2006MNRAS.372L..33B,2006ApJ...653L.125P}. \citet{LP00} gave the criterion for distinguishing the thin channel and the thick channel. For thick channel:
\begin{equation}
\label{eq1}
\Delta v^2>\delta v^2
\end{equation}
Where $v$ is the velocity component along the line of sight, $\Delta v$ is the velocity channel width, $\delta v$ is the velocity dispersion. The criterion to identify thick channel given in \citet{LP00} is a lower limit. The data which contains no channel but accumulates intensity information along the line-of-sight (LOS) automatically meets with the thick channel criterion, for example, the \ion{H}{1} column density data and dust emission data. Hence, we expect the Intensity Gradients Technique proposed in this work (see Sec.\ref{sec.method}) is applicable to those data.
\subsection{Properties of velocities and densities in MHD simulations}
\label{sec.2.3}
\begin{figure}[t]
	\centering
	\includegraphics[width=0.99\linewidth,height=0.55\linewidth]{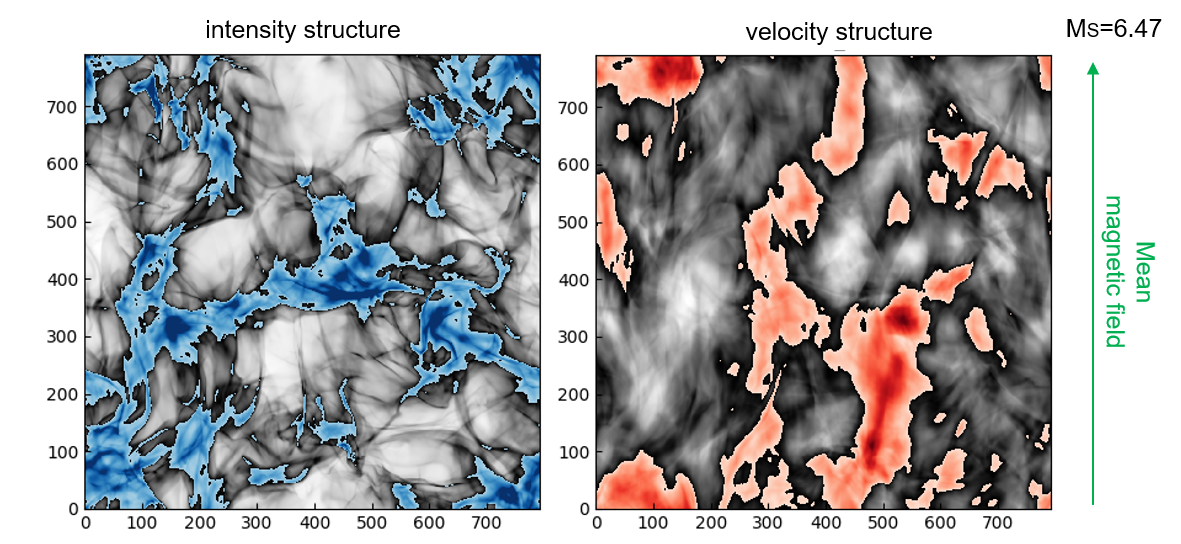}
	\caption{\label{fig:vis} An example of structures in numerical intensity map and velocity map (M$_s$= 6.47). The colorful isocontours (blue and red) correspond to the region in which its intensity (left)/velocity (right) is larger than 75th percentile of the full map.}
\end{figure}
Density and velocity fields, in general, have different statistics and contain different information. Therefore, velocity and density gradients can behave differently, e.g., in self-gravitating regions and shocks. As shown in Fig.~\ref{fig:vis}, high contrast density structures (highlighted by blue color) in intensity map are perpendicular to the magnetic field, but low contrast density structures (grey color) are parallel to the magnetic field. These clumpy dense structures which exhibit scale-dependent anisotropy are earlier seen in \citet{2005ApJ...624L..93B}. They studied moderately magnetized  media ($\beta\sim 1$) and found that $E\sim k^{-5/3}$. Later \citep{2019arXiv190506341X} explained that perpendicular turbulent mixing of density fluctuations entails elongated low-density structures aligned with the local magnetic field, while high-density filaments compressed by shock are perpendicular to the local magnetic field. However, for velocity structures, they are always parallel to local magnetic fields.
	
\begin{figure}[t]
	\centering
	\includegraphics[width=0.99\linewidth,height=0.9\linewidth]{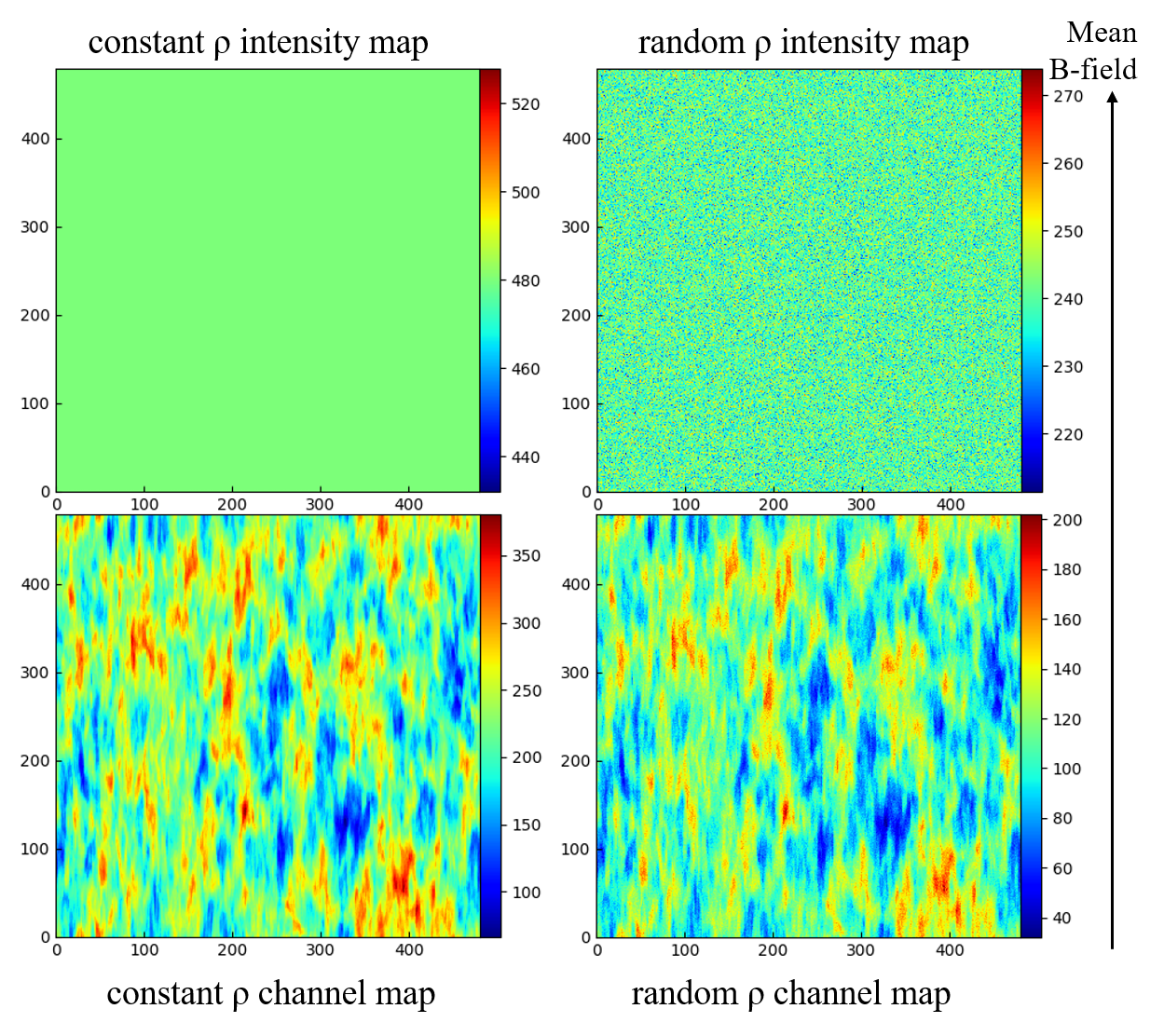}
	\caption{\label{fig:constant} An illustration of the difference between thick channels (denoted as intensity maps, top panel) and thin channel maps (denoted as channel maps, bottom panel) in PPV cube. We produce the PPV cubes from density PPP cube using: (i) constant density field (left column) (ii) random density field (right column), while keeping the original velocity field un-changed ($M_s$=20).}
\end{figure}
	
Furthermore, one can study the variation of gradients as the relative contribution of density and velocity changes by varying the thickness of velocity channels. In Fig.~\ref{fig:constant}, we give an illustration of the difference between thick channels (denoted as intensity map) and thin channels (denoted as channel map) using synthetic observation data. We produce the PPV cubes from density PPP cube using: (i) uniform density distribution (ii) Gaussian random density distribution while keeping the original velocity field un-changed ($M_s$=20). We can see that for both different density distributions, their thin channel maps show similar structures, while their thick channel maps (i.e., intensity map) are different. Those similar structures in the thin channel are created by velocity caustics, i.e., they are not practical density structures. Therefore, thin channel map contains more information of velocity field rather than density field, but the opposite for intensity maps.
	
	
\section{MHD simulation data}
\label{Sec.MHD data}
The numerical 3D MHD simulations are generated by \textbf{ZEUS-MP/3D code \citep{2006ApJS..165..188H}}, which uses a single fluid, operator-split, staggered grid MHD Eulerian assumption. 
	
To emulate a part of interstellar cloud, periodic boundary conditions and solenoidal turbulence injections are applied in our simulations. We employ various Alfvenic Mach numbers M$_{A}=\frac{v_{L}}{v_{A}}$ and sonic Mach numbers  M$_{s}=\frac{v_{L}}{v_{s}}$ in our simulations, where $v_{L}$ is the injection velocity, while $v_{A}$ and $v_{s}$ are the Alfvenic and sonic velocities respectively (See Table \ref{tab:sim} for details). In the simulations, we also employ various compressibility $\frac{\beta}{2}=(\frac{v_{s}}{v_{A}})^2$ of MHD turbulence. The plasma shows low compressibility $\frac{\beta}{2}<1$ when the magnetic pressure of  plasma is larger than the thermal pressure (i.e.M$_{A}<$M$_{s}$, high magnetization level), while the domain M$_{A}>$M$_{s}$ corresponds to the pressure dominated plasma with $\frac{\beta}{2}>1$. 
	
Furthermore, we refer to the simulations Table \ref{tab:sim} by their model name. For instance, our figures will have the model name indicating which data cube was used to plot the figure. The ranges of M$_{s}$ and M$_{A}$ are selected so that they cover different possible scenarios of astrophysical turbulence from very subsonic to supersonic cases. We considering an ideal case without self-gravity, expect for simulation M$_s$0.2. The data has been used  to set up a three-dimensional, uniform, isothermal turbulent medium \citep{YL17a,YL17b,2018ApJ...865...46L,2018ApJ...865...54Y,2019MNRAS.tmp.1175Z}.
	
\begin{table}
	\centering
	\label{tab:sim}
	\begin{tabular}{| c | c | c | c | c |}
	\hline
	Set & Model & M$_s$ & M$_A$  & resolution \\ \hline \hline
		&M$_A$0.2 & 7.31 & 0.22 & $792^3$\\  
		&M$_A$0.4 & 6.10 & 0.42 & $792^3$\\
		&M$_A$0.6 & 6.47 & 0.61 & $792^3$\\  
		&M$_A$0.8 & 6.14 & 0.82 & $792^3$\\  
	   A&M$_A$1.0 & 6.03 & 1.01 & $792^3$\\
		&M$_A$1.2 & 6.08 & 1.19 & $792^3$\\
		&M$_A$1.4 & 6.24 & 1.38 & $792^3$\\
		&M$_A$1.6 & 5.94 & 1.55 & $792^3$\\
		&M$_A$1.8 & 5.80 & 1.67 & $792^3$\\
		&M$_A$2.0 & 5.55 & 1.71 & $792^3$\\\hline
		&M$_s$0.2 & 0.2 & 0.02 & $480^3$\\
		B&M$_s$20.0 & 20.0 & 0.2 & $480^3$\\
		\hline
		\end{tabular}
		\caption{Description of our MHD simulations. M$_s$ and M$_A$ are the instantaneous values at each the snapshots are taken. Three-dimensional, uniform, and isothermal turbulent medium.}
\end{table}
\section{Methodology}
\label{sec.method}
\subsection{The Intensity Gradients Technique}
The velocity gradients in thin velocity channels and the intensity gradients in thick velocity channels are obtained using analytical description of Position-Position-Velocity (PPV) \citep{LP00,2008ApJ...686..350L} cubes. The thin velocity channel map \textbf{Ch(x,y)}, in which the velocity contribution in velocity channels dominates over the density contribution, is calculated as:
\begin{equation}
Ch(x,y)=\int_{v_0-\Delta v/2}^{v_0+\Delta v/2}\rho(x,y,v) dv
\label{eq1}
\end{equation}
where $\rho$ is gas density, $v$ is the velocity component along the line of sight, $\Delta v$ is the velocity channel width satisfied with Eq.~\ref{eq1}, $v_0$ is the velocity corresponding to the peak position in PPV's velocity profile. As for the intensity map \textbf{I(x,y)} in which density contribution is dominated is produced by doing integral along the velocity axis of PPV cube:
\begin{equation}
I(x,y)=\int \rho(x,y,v)dv
\label{eq2}
\end{equation}
	
We note that in the case of sub-sonic turbulence, it is advantageous to use velocity centroids to reveal velocity statistics \citep{2017MNRAS.464.3617K,2005ApJ...631..320E}. However, the potential disadvantage of traditional centroids is that the entire spectral line is used while in some cases different parts of spectral line reflect magnetic fields in spatially different regions. This is the case, for instance, for the \ion{H}{1} measurements where the galactic rotation curve cannot isolate a particular region of the galaxy. We therefore define the "reduced velocity centroid" map \textbf{R(x,y)},  which make use of part of spectral line only, as:
\begin{equation}
R(x,y)=\int_{v_0-\Delta v/2}^{v_0+\Delta v/2} \rho(x,y,v)vdv 
\label{eq4}
\end{equation}
	
In this work, we denote the gradients calculated from \textbf{I(x,y)} as Intensity Gradients (IGs), while from \textbf{Ch(x,y)} as Velocity Channel Gradients (VChGs). The study and application of \textbf{R(x,y)}, i.e., Reduced Velocity Centroid Gradients (RVCGs), is available in \citet{2019ApJ...874...25G} and \citet{2018ApJ...853...96L}. The gradient angle at each pixel $(x_{i},y_{j})$ in the plane-of-the-sky (POS) is defined as:
\begin{equation}
\bigtriangledown_{i,j}f=tan^{-1}[\frac{f(x_{i},y_{j+1})-f(x_{i},y_{j})}{f(x_{i+1},y_{j})-f(x_{i},y_{j})}]
\end{equation}
$f(x,y)$ can be \textbf{I(x,y)}, \textbf{Ch(x,y)}, or \textbf{R(x,y)}. After the pixelized gradient field is established, the sub-block averaging method is then applied to predict the direction of magnetic fields through gradients in a statistical region of interest \citep{YL17a}. The use of sub-block averaging comes from the fact that the orientation of turbulent eddies with respect to the local magnetic field is a statistical concept. When statistical samples are sufficiently large, the histogram of gradient orientations will show a Gaussian profile \citep{YL17a}. Within a sub-block, we obtained the most probable orientation, which is the peak of the Gaussian corresponding to the local direction of the magnetic field within the block.
	
The correspondence of gradients rotated by $90^\circ$ and magnetic fields is quantified using the \textbf{Alignment Measure} (AM): $AM=2(\langle cos^{2} \theta\rangle-\frac{1}{2})$, where $\theta$ is the relative angle between rotated gradients and orientations of magnetic fields. If the two measures provide identical results, AM = 1, while AM = -1 indicates the relative angle is $90^\circ$.

\subsection{Shock identification  algorithm}
The sonic Mach number M$_s$, which is the ratio of the turbulent injection velocity and the speed of sound, characterizes the compressibility of turbulent flow.  When M$_s$ gets large, i.e., M$_s>1$, supersonic flows will inevitably form shock waves due to stronger compression. Shocks are a vital process, for instance, MHD simulations by \citet{1998ApJ...508L..99S} found that 50\% of turbulent energy is dissipated to shocks and the properties of turbulent gas are significantly modified. As shown in Fig.~\ref{fig:constant}, density fluctuation dominates over velocity fluctuation in the thick channel map, i.e., the gas structure in super-sonic turbulence is a practical density structure i.e., the formation of shocks. Therefore, it is potentially possible to identify the shock structure in PPV cubes using thick channel map. In this work, we propose a new algorithm to identify shock structures using IGs, as well as study how shocks change the alignment of gradient vectors and the underlying magnetic field. 
	
To start with, we focus on removing the strong J-shocks \citep{2009ASPC..414..453D}. For Jump discontinuity, the change of density across shocks is very significant compared to surrounding environments. Therefore a higher density gradient amplitude is found across the shock front. Hence we sort out the intensity map according to their amplitudes. Denote the gradients amplitude x, and the global mean $\mu$, the standard deviation $\sigma$, the Z-score of x is defined as \citep{YL17b}
\begin{equation}
Z(x) = \frac{x -\mu}{\sigma}
\end{equation}
	
A higher positive Z score stands for regions with gradient amplitude above the system average. Since areas with higher amplitude correspond to those with J-shocks, the structure with positive Z-scores is identified as a candidate of shocks.

\section{Numerical simulation results}
\label{sec:simulation}
	
\begin{figure}[t]
	\centering
	\includegraphics[width=0.99\linewidth,height=1.5\linewidth]{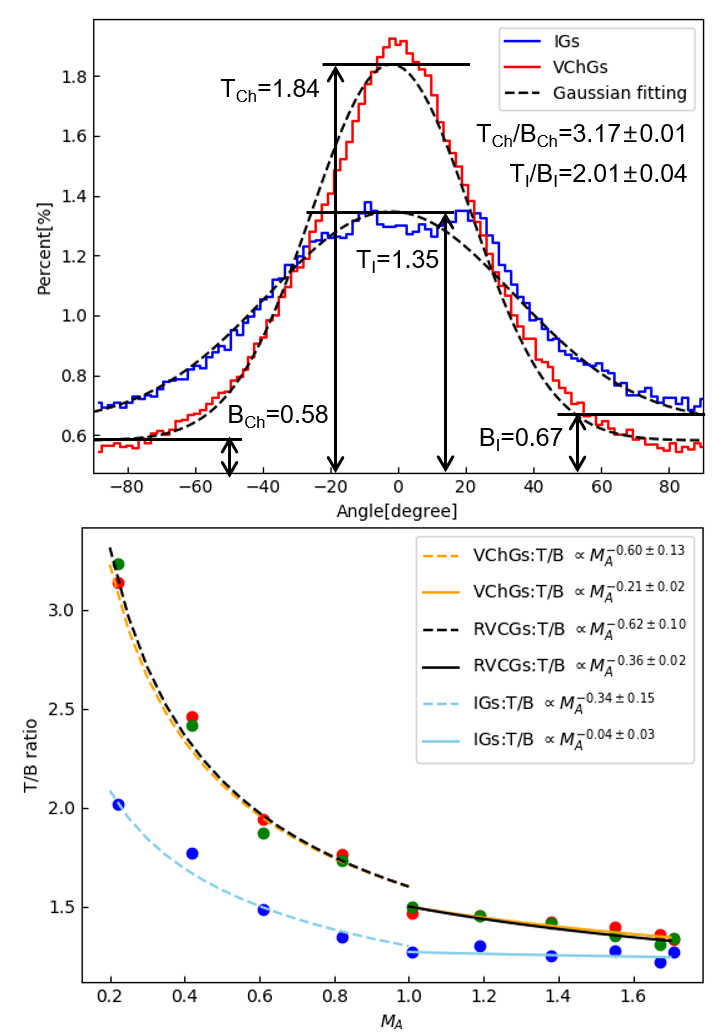}
	\caption{\label{fig:TB_hist} \textbf{Top:} the histogram of gradients orientation. T denotes the maximum value of the fitted histogram, while B is the minimum value. \textbf{Bottom:} a plot shows the correlation between T/B ratio and Alfven Mach number M$_A$.}
\end{figure}


\subsection{Properties of gradient distributions}
\begin{figure*}[t]
	\centering
	\includegraphics[width=.99\linewidth,height=0.58\linewidth]{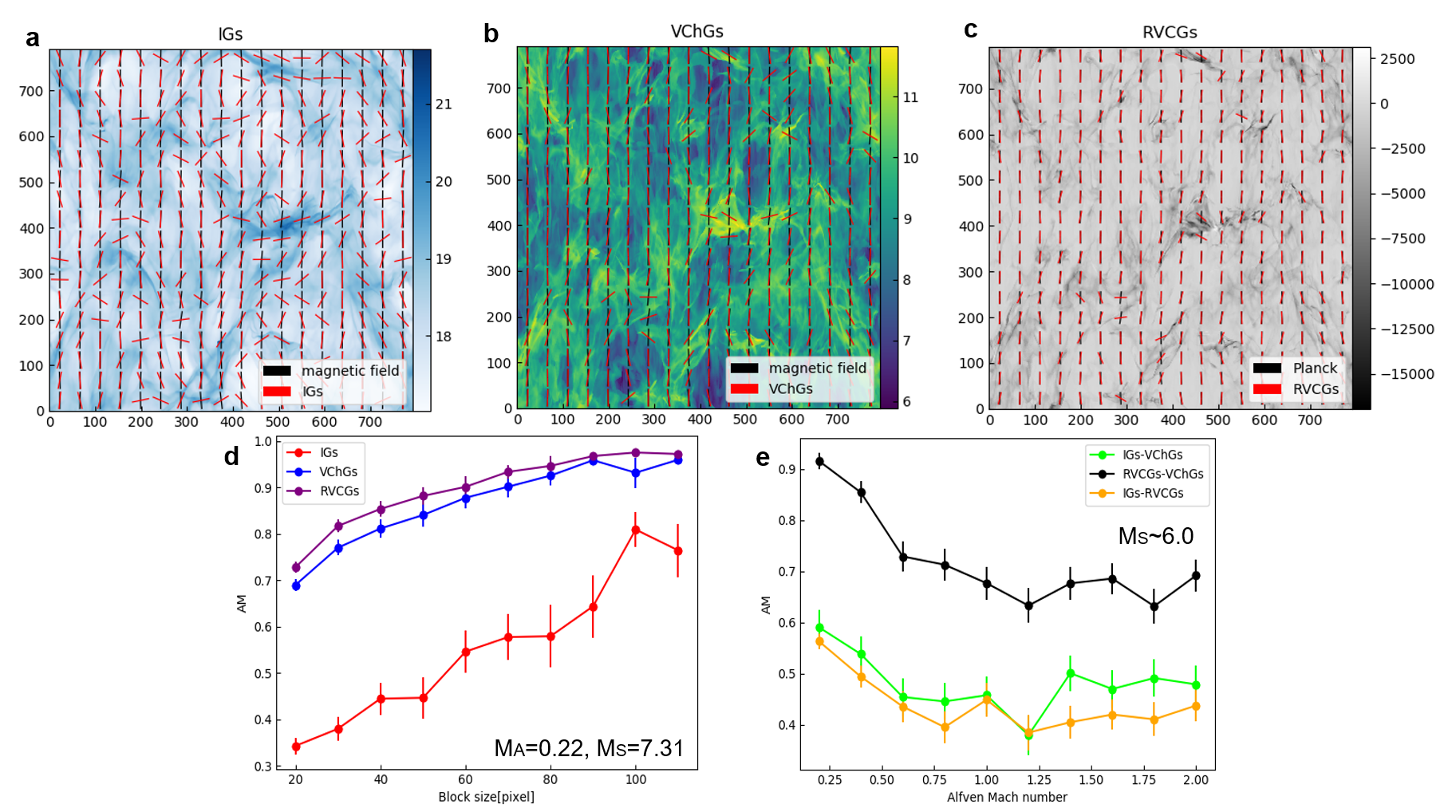}
	\caption{\label{fig:IGs} \textbf{a, b, c:} an example of the magnetic fields inferred from IGs, VChGs, and RVCGs respectively with sub-block averaging applied, using simulation M$_A$0.2. \textbf{d:} an example of the correlation between AM (gradients and polarization) and the block size, using simulation M$_A$0.2. \textbf{e:} the variation of the alignment between each gradients type (IGs, VChGs, RVCGs) as a function of M$_A$.}
\end{figure*}
	
VGT has been introduced in \citet{2018ApJ...865...59L} to obtain a reliable estimation of the magnetization of the media in both \ion{H}{1} data  \citep{2018ApJ...865...59L} and molecular clouds \citep{survey}. As shown in \citet{2018ApJ...865...46L}, one could estimate the magnetization through the power-law correlation of Alfvenic Mach number M$_A$ and the dispersion of the velocity gradient distribution. The distribution of velocity gradients orientation is generally Gaussian. For high magnetization, the distribution is sharply peaked, and the dispersion is small. This corresponds to the excellent alignment of individual gradient vectors and the magnetic field direction. For low magnetized media, the dispersion increases. As for intensity gradients, its distribution shows similarly behaviors as velocity gradients. Fig.~\ref{fig:TB_hist} shows an example of normalized histogram of gradients orientation without sub-block averaging. We see that the distributions of both IGs and VChGS orientations is Gaussian, while IGs is more dispersed than VChGs. The uncertainty of T/B ratio is negligible. This difference can be explained by the presence of shocks in intensity map. The intensity gradients are gradually changing their direction to be parallel to magnetic fields when getting close to the shock front. In this case, the distribution of intensity gradients orientation gets more dispersion than the one of velocity gradients orientation.
	
We quantify the gradients dispersion by T/B ratio, where T denotes the maximum value of the fitted histogram of gradients orientation, while B is the minimum value (see Fig.~\ref{fig:TB_hist}). The uncertainty is given by the error of Gaussian fitting within 95\% confidential level. Fig.~\ref{fig:TB_hist} shows the correlation of T/B ratio and M$_A$. We find that generally, T/B ratio decreases with the increment of M$_A$. While there is a well-fit power law of $T/B\propto M_A^{-0.60\pm0.13}$ for VChGs, $T/B\propto M_A^{-0.21\pm0.02}$ for IGs, and $T/B\propto M_A^{-0.62\pm0.10}$ for RVCGs in the case of sub-Alfvenic turbulence $M_A<1$, while $T/B\propto M_A^{-0.36\pm0.02}$ for VChGs, $T/B\propto M_A^{-0.04\pm0.03}$ for IGs, and $T/B\propto M_A^{-0.36\pm0.02}$ for RVCGs when M$_A>1$.  Our results coincide with the results of velocity centroid gradients in \citet{2018ApJ...865...46L}. They shows $T/B\propto M_A^{-0.46\pm0.18}$ for velocity centroid gradients in case of M$_A<1$. The change in power-law index for M$_A>1$ is expected, as discussed in \citet{2018ApJ...865...46L}, the nature of turbulence changes when the injection velocity becomes higher than the Alfven speed. In this situation, the large-scale motions of eddies are dominated by hydro-type turbulence, and the directions of magnetic fields within flows are significantly randomized. This changes the distribution function of gradient orientations. In addition to the well-fit power law for velocity gradients, including VChGs, RVCGs, and VCGs, IGs also shows corresponding reaction with respect to the variation of magnetization. IGs therefore as a complementary tool can be used synergetically with VGT for estimating the magnetization level. 

\subsection{Tracing magnetic field morphology}
From what we have discussed in \S~\ref{sec:theory}, it follows that the correlation with magnetic fields is expected not only for velocity gradients, but also intensity gradients. In \citet{GL17}, the relative orientation between intensity gradients and magnetic fields has been primarily explored. It was shown that raw intensity gradients (without the sub-blocked averaging method applied) are not well correlated with the direction of magnetic fields, giving much larger error estimates for the direction of magnetic fields. Hence, we go further by applying the sub-block averaging method \footnote{The sub-block averaging method was initially developed for VGT \citep{YL17a}. It is not just a smoothing method
for suppressing noise in a region, but used to increase the reliability of important statistical measurements.} to IGs, in order to have a reliable determination of both the direction of IGs and the statistical significance of this determination. We expect IGs would be a complementary tool to VGT in terms of tracing magnetic fields and getting additional information about shocks. 
	
We show an example of IGs and VChGs using the simulation M$_A$0.2. Fig.~\ref{fig:IGs} shows 2D vector maps of magnetic fields traced by IGs, VChGs, and RVCGs with sub-block size 44 pixels. RVCGs shows a better alignment (AM=0.87$\pm0.02$) with the magnetic field than VChGs (AM=0.82$\pm0.02$) and IGs (AM=0.47$\pm0.03$). The uncertainty is given by the standard error of the mean, that is, the standard deviation divided by the square root of the sample size.
\begin{figure}[t]
	\centering
	{\includegraphics[width=.99\linewidth,height=0.5\linewidth]{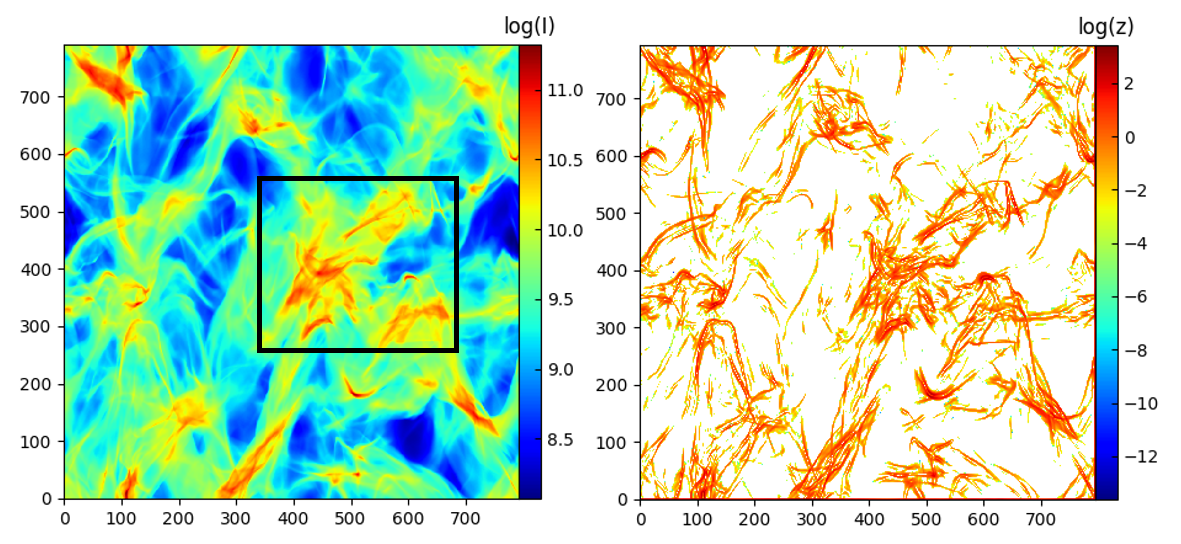}}
	\caption{\label{fig:shocks} The figure shows how the intensity gradient amplitude with positive Z-score (left) is related to high density shock structure (right). We use the simulation M$_A$0.4 here.  The black box is indicating a zoom-in region for Fig.~\ref{fig:serpens}.}
\end{figure}

We further explore the ability of IGs, RVCGs, and VChGs in tracing magnetic fields, in terms of the sub-block size, which corresponds to the measurement scale in observation. Fig.~\ref{fig:IGs} shows the Alignment Measure (AM, between rotated gradients and magnetic fields) as a function of the block size using simulation M$_A$0.2. We find that AM of IGs, RVCGs, and VChGs are positively proportional to the sub-block size. Since a large sub-block contains more samples points, the statistical result is more accurate. In addition, RVCGs and VChGs generally give better alignment than IGs and at small scale (block size), RVCGs and VChGs still show good alignment (AM$\sim 0.70\pm0.01$). As our theoretical expectation, the velocity fluctuation follow the same GS95 anisotropy relation in all scale, while it is not the case for density fluctuation.

We see plot the alignment between different gradients as a function of M$_A$ and M$_S$ in Fig.~\ref{fig:IGs}, keeping block size=44 pixels. We see that for sub-Alfvenic turbulence, the alignment between different gradients is decreasing, while for super-Alfvenic AM tends to be stable. The change in trend is similar to the one obtained from gradients distributions, since the nature of turbulence changes when its kinematic property becomes more important than the magnetic filed. In any case, VChGs and RVCGs are well aligned with each other, which demonstrate that VChGs contains the information of velocity field. Importantly, the correlation therefore potentially provides the possibility to measure M$_A$ using the alignment between different gradients in future studies.
	
To summary, for the sub-sonic case, velocity centroid is a better way in representing velocity statistics \citep{2017MNRAS.464.3617K,2005ApJ...631..320E}. In the fact that the center of spectral line is saturated due to absorption effects, it is good to use only the informative part of the line. Therefore, we propose to apply VChGs to trace the magnetic field orientation in super-sonic turbulence, while RVCGs for sub-sonic turbulence. As for IGs, it can be used as a complementary tool when velocity information is not available, for example, the \ion{H}{1} column density data.


\subsection{Identify shock structures}
Highly contrast density structures (i.e., shocks) are perpendicular to magnetic fields, but lowly contrast density structures are parallel to magnetic fields, while this is not the case for velocity structure\footnote{ \citet{2019arXiv190403173Y} shows that gradients of intensity in thick channels get perpendicular to gradients of intensity in thin channels testify in favor of thin channels of representing velocity fluctuations. } (see Sec.~\ref{sec.2.3}). Insight of this difference, we consider that the higher contrast shock is one possible obstacle for intensity gradients in term of tracing magnetic fields\footnote{ Note that thin velocity channels are mixtures of densities and velocities, while the contribution from velocities is dominating over densities \citep{LP00}. Therefore, although insignificant, the performance of gradients in thin velocity channels is also affected by densities.}. 
\begin{figure}[t]
\centering
\includegraphics[width=0.98\linewidth,height=0.85\linewidth]{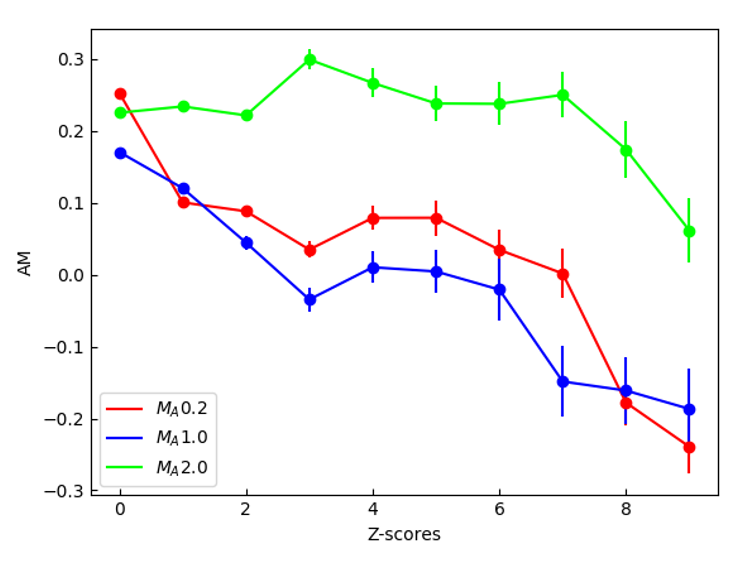}
\caption{\label{fig:IG_shocks} The correlation of AM and Z-scores. Each AM is calculated from raw gradients without sub-block averaging and magnetic fields corresponding to same Z-score, but not the overall AM.} 
\end{figure}

\begin{figure*}
\centering
\includegraphics[width=1.0\linewidth,height=0.81\linewidth]{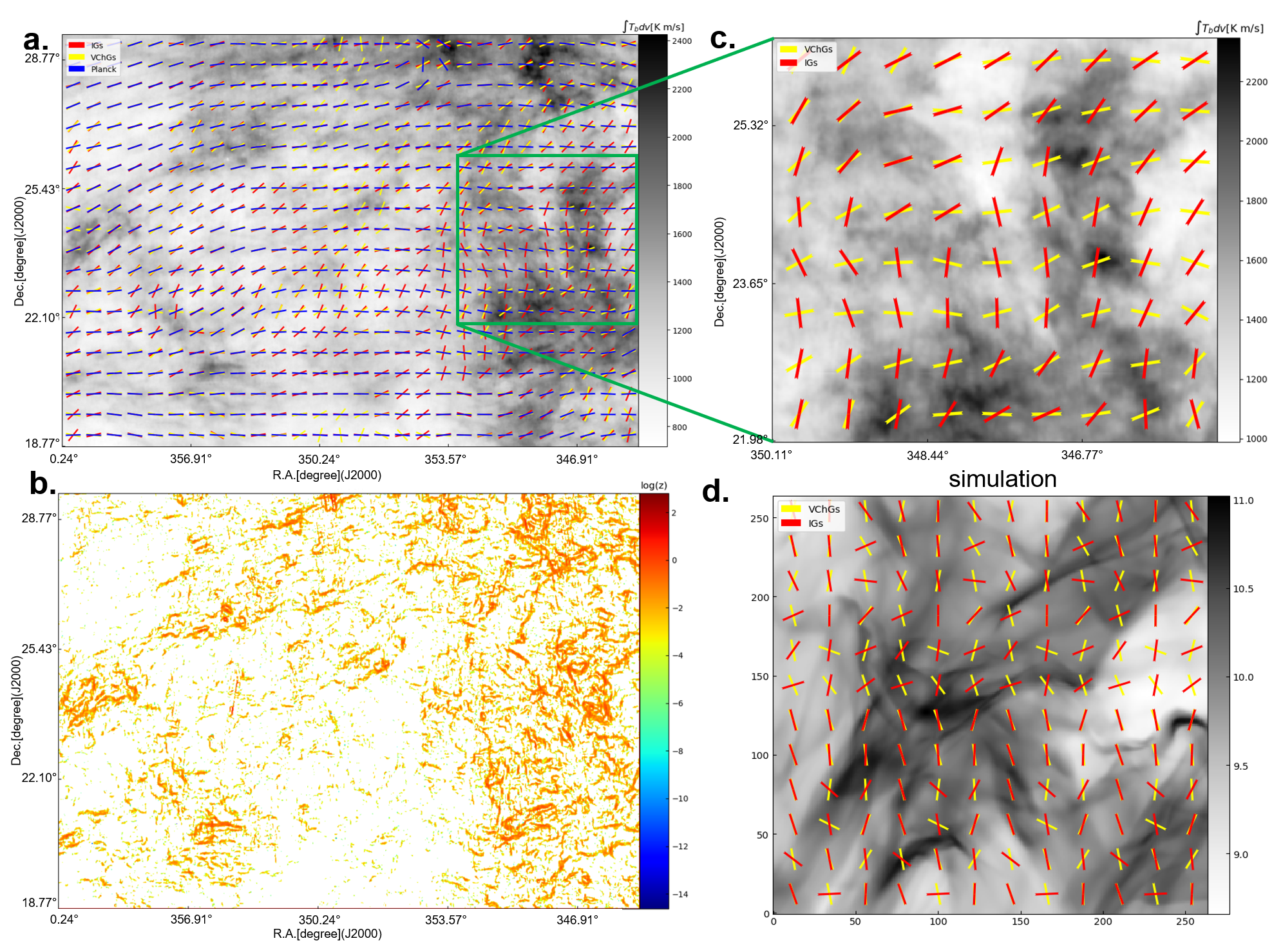}
\caption{\label{fig:serpens} \textbf{Panel a:} The magnetic field morphology predicted by Planck 353GHz polarization (blue segments), VChGs (yellow segments) and IGs (red segments). The sub-block size is selected as 32 pixels (i.e. effective resolution $\sim 0.5^\circ$). \textbf{Panel b:} the shock structure identified using the Z-score algorithm. \textbf{Panel c:} a zoom-in region which shows that IGs and VChGs get to be perpendicular. \textbf{Panel d:} a zoom-in region from our numerical simulation shown in Fig.~\ref{fig:shocks}.}
\end{figure*}
	
Fig.~\ref{fig:shocks} shows how the shock identification algorithm works on thick channel map. We sort out the intensity map according to their Z-score and wash out the one with negative Z-score. We see the intensity gradient amplitude with positive Z-score is well correlated to high density shock structure. In addition, we sort out IGs according to Z-scores and calculate their AM without applying the sub-block averaging method. Fig.~\ref{fig:IG_shocks} shows the plot of AM as a function of the Z-scores. We see that the alignment is decreasing with higher Z-scores. However, the alignment changes rapidly in case of a strong magnetic field (low M$_A$). When Z-score is more significant than 8, the alignment of M$_A$0.2 and M$_A$1.0 is approximately -0.2, while it is still positive for M$_A$2.0. As explained in \cite{2019arXiv190506341X}, high-density filaments compressed by shock are perpendicular to the local magnetic field.  We, therefore, expect that the shock can be identified with Z-scores larger than 8 in case of sub-Alfvenic turbulence. Besides, \cite{Soler2013} concluded that magnetic fields and intensity gradients get parallel when self-gravity is dominating over turbulence. However, our results show that for high M$_s$ turbulence, we can have the change of the relative orientations even without self-gravity.
	

\section{Observational Results}
\label{sec.obs} 
To demonstrate the observational applicability of IGT in tracing magnetic filed, and identifying shock structure with the newly developed algorithm, we apply the technique to GALFA-\ion{H}{1} spectroscopic data \citep{2018ApJS..234....2P}. The data selected from GALFA-\ion{H}{1} survey corresponds to the region stretches from R.A.$\sim 345.35^\circ$ to R.A. $\sim 0.24^\circ$.  We analysis the \ion{H}{1} data with velocity range from -21$km/s$ to 21$km/s$. Magnetic field orientation is derived using Planck Collaboration III 2018 PR3 353 GHz polarization data \citep{2018arXiv180706207P}, where the signal-to-noise ratio (S/N) of dust emission is maximum, as a tracer of the magnetic field \footnote{Planck is a project of the European Space Agency (ESA) with instruments provided by two scientific consortia funded by ESA member states (in particular the lead countries France and Italy), with contributions from NASA (USA) and telescope reflectors provided by a collaboration between ESA and a scientific consortium led and funded by Denmark.}. The polarization angle $\phi$, and POS magnetic field orientation angle $\theta_B$ can be derived from the Stokes I, Q, and U parameter using the relation:
\begin{equation}
\begin{aligned}
\phi&=\frac{1}{2}*arctan2(-U,Q)\\
\theta_B&=\phi+\pi/2
\end{aligned}
\end{equation}
	
The minus sign of U converts the Planck data to IAU convention, where the polarization angle is counted positively from the Galactic north to the east. Before calculating $\phi$, one should carefully transform the Stokes U, Q maps from Galactic coordinate to Equilateral coordinate. As for the calculation of gradients, we implement the sub-block averaging method with block size equals 32 pixels ($\sim 0.5^\circ$) and the moving window method to IGs with width equals 2 following \citet{survey}. 
	
\begin{figure}
	\centering
	{\includegraphics[width=.99\linewidth,height=2.15\linewidth]{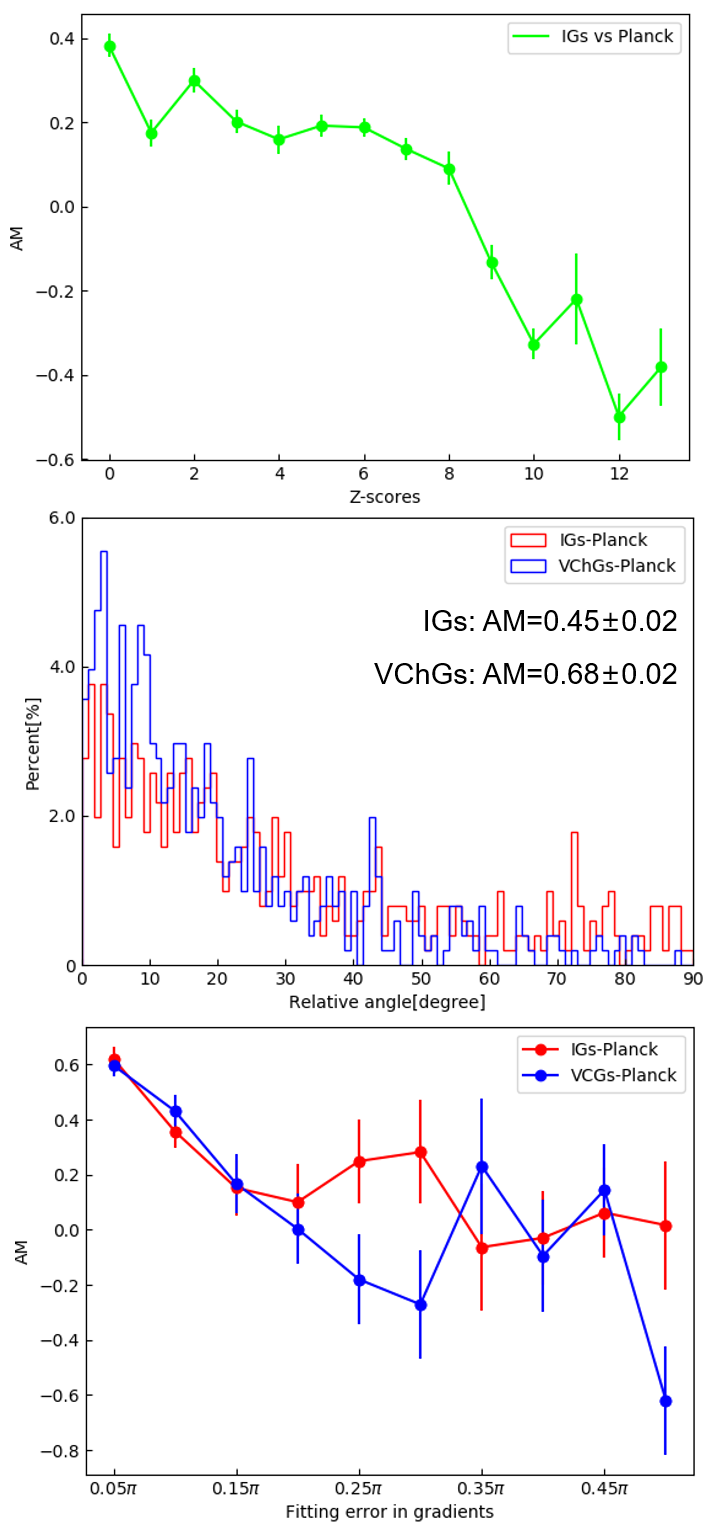}}
	\caption{\label{fig:DM_serpens} \textbf{Top panel:} plot of AM as a function of Z-scores using raw gradients without sub-block averaging. AM=0.0 indicates random distribution, i.e. neither parallel nor perpendicular. \textbf{Middle panel:} the histogram of the relative angle between the magnetic field inferred from IGs/VChGs and Planck polarization. \textbf{Bottom panel:} The variation of AM with respect to the fitting error in gradients. The error is from fitting the histogram of gradients orientation within a sub-block.}
\end{figure}

Fig.~\ref{fig:serpens} shows the $B_{POS}$ morphology inferred from IGs, VChGs, and Planck polarization, as well the shock structure identified by IGT. Visually we see that VChGs (AM=$0.68\pm0.02$) aligns with the magnetic field inferred from Planck polarization better than IGs (AM=$0.45\pm0.02$). However, there is a significant dis-alignment of IGs and VChGs in the upper part of Fig.~\ref{fig:serpens}a where we find a lot of shocks. As for the deviation between the gradients and the magnetic fields inferred from Planck polarization, the error from fitting the histogram of gradients orientation within a sub-block is one possible contribution. In Fig.~\ref{fig:DM_serpens}, we plot the variation of AM with respect to the fitting error in gradients. We bin the fitting error into ten uniform intervals from 0 to $\pi/2$ and take the average value of AM in each interval. We see that AM is generally decreasing with the increment of fitting error. The deviation is, therefore, possibly from the fitting uncertainties.

Also, we plot the histogram of the relative angle between the rotated IGs/VChGs and the magnetic field inferred from Planck polarization in Fig.~\ref{fig:DM_serpens} middle panel. For both IGs and VChGs, the histogram is concentrated on $\sim$5$^{\circ}$, with AM=0.68$\pm$0.02 for VChGs and AM=0.45$\pm$0.02 for IGs. It is indicates that VChGs is more reliable and accurate than IGs in terms of magnetic field tracing, by comparing with the Planck polarization.

Furthermore, we study the relative orientation between VChGs and IGs. Fig.~\ref{fig:serpens} observationally and numerically shows VChGs and IGs in a zoom-in region which is full of shocks in terms of our analysis. We see that IGs and VChGs become perpendicularly aligned. As we illustrated in Fig.~\ref{fig:vis}, high contrast density structures compressed by shock shows different orientation with respect to velocity structures. It, therefore, confirms LP00's theory that the thick velocity channel and thin velocity channel contains various information, i.e., the contribution from velocity field is dominating in narrow velocity channels. 
	
The top panel in Fig.~\ref{fig:DM_serpens} shows the variation of raw gradients without sub-block averaging with respect to Z-score. Each AM value is calculated from raw gradients and magnetic fields corresponding to the same Z-score, but not the overall AM. We see that AM is negatively proportional to Z-score, which indicates that the intensity gradient tends to be parallel to magnetic fields in front of shocks. This coincides with our numerical simulation results and theoretical consideration. We therefore expect the structure with Z-score larger than 10 can be identify as shocks and we propose to re-rotate raw intensity gradients in front of shocks by 90$^\circ$ again before implementing the sub-block averaging method.
	
\section{Histogram of the Relative Orientation}
\subsection{Differences and comparison with HRO}
The Histogram of Relative Orientations (HRO) technique was introduced by \cite{Soler2013}, which empirically used a relative orientation angle $\theta$ between the density gradient and the magnetic field in each pixel to characterize the direction of column density structures in a histogram form. \textbf{IGT is different from HRO technique, which requires polarimetry data to define the direction of magnetic fields. IGT is polarization-independent and is the way of finding magnetic field direction, using the sub-block averaging method.}  The comparison of IGT and HRO is summarized in Tab.~\ref{tab:HRO}. 

\begin{table}
\centering
\label{tab:HRO}
\begin{tabular}{| c | c | c |}
\hline
Technique & IGT & HRO \\ \hline \hline
Necessity of polarimetry & No & Yes \\  \hline
Trace magnetic field & Yes (limited) & No \\\hline
Identify gravitational collapse &  Yes (with VGT) & No \\\hline
Trace shocks & Yes (with VGT) & No \\\hline
Measure M$_A$ & Yes (limited) & No \\\hline
Synergy with VGT & Yes & Yes \\
\hline
\end{tabular}
\caption{Comparison of IGT and HRO.}
\end{table}
\cite{Soler2018} tried to improve their technique, called the Histogram of Oriented Gradients (HOG), which uses gradients in thin velocity channels to compare systematically the gradients contours that might be common \ion{H}{1} and $^{13}$CO emission. However, \cite{Soler2018} consider the gradients in thin velocity channels as pure intensity gradients. \cite{Soler2018} disregarded the effect of forming intensity fluctuations by the velocity crowding effect \citep{LP00,LP04,LP06,LP08,KLP16,KLP17b,2006ApJS..165..512K,2006AIPC..874..301L,2006MNRAS.372L..33B,2006ApJ...653L.125P}, which is clearly demonstrated in Fig.~\ref{fig:constant}. They therefore erroneously assumed that in thin velocity channels all intensity fluctuations are due to density enhancements. \cite{Soler2018} and \citet{2018ApJ...865...59L} used different approaches in the analysis of intensity distributions in thin velocity channels\footnote{The intensity fluctuations in thin velocity channels is the basis of VChGs technique \citep{2018ApJ...865...59L}.}. The former does not use the procedure of sub-block averaging, which is disadvantageous and prevents \cite{Soler2018} from reliably tracing magnetic field direction as it is demonstrated in \citet{2018ApJ...865...59L,survey,Laura,2019ApJ...874...25G}
	
To compare IGT and HRO, we follow the recipe used in \citet{Soler2013}, which calculates $\theta$ using a combination of the scalar and vector product of vectors:

\begin{equation}
\theta=arctan(\frac{|\textbf{B}\times\nabla I|}{\textbf{B}\cdot \nabla I})
\end{equation}
where \textbf{B} is the magnetic field, while I is the intensity for spectroscopic data. To quantify the progressive change of relative orientation in a histogram form (e.g., the HRO curve changes from convex to concave). \citet{Soler2013} defined the histogram shape parameter: $\zeta=A_c-A_e$, where and $A_c$ is the area under the central region of the HRO curve ($-0.25<cos\theta<0.25$), $A_e$ is the area in the extremes of HRO ($-1.0<cos\theta<-0.75$ and $0.75<cos\theta<1.0$). This parameter characterizes a curve peaking at $cos\theta\sim0$ (convex) as $\zeta>0$ whereas a curve peaking at $cos\theta\sim \pm 1.0$ (concave) corresponds to $\zeta<0$ and a flat distribution corresponds to $\zeta\sim0$. The uncertainty in the determination of $\zeta$ is given by the standard deviation around the calculated area in each region.
\begin{figure}[t]
\centering
\includegraphics[width=.99\linewidth,height=.85\linewidth]{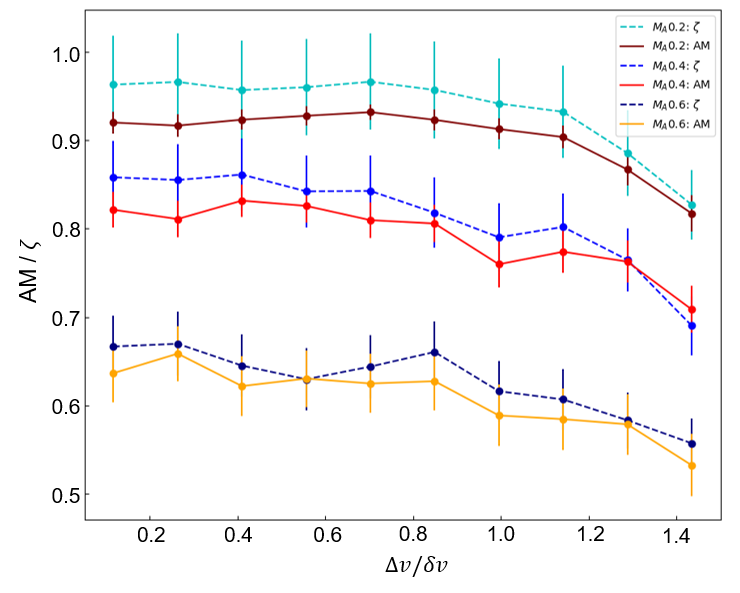}
\caption{\label{fig:HRO} The correlation between $\Delta v/\delta v$ (x-axis) and AM/$\zeta$ (y-axis), where $\Delta v$ is the velocity channel width, $\delta v$ is the velocity dispersion. The dash line indicates $\zeta$ used in HRO, while the solid line means AM used in gradients technique. }
\end{figure}

We compare $\zeta$ with our AM used in gradients technique concerning the performance in analyzing the relative orientation between density gradients and magnetic fields. Insight of the fact that one can study the relative contribution of density and velocity by varying the thickness of the slice, we also extend the HRO analysis to velocity gradients. Fig.~\ref{fig:HRO} shows the correlation of AM/$\zeta$ and the width of the channel. We see that the alignment between gradients and magnetic fields is decreasing for a thicker channel, which coincident with our theoretical consideration. We find that both HRO analysis and our AM analysis give similar results, whereas HRO usually shows a larger value. Although HRO is developed initially for analyzing density gradients, we show that it is also applicable to velocity gradients in thin channels. 

\subsection{Modifications to HRO}
\label{subsec:HRO}
\cite{Soler2013} was using the histogram of $cos (\theta$) weighted by $|\nabla I|$ to characterize the relative orientation between density gradients $\nabla I$ and $B_{POS}$\footnote{In \cite{Soler2013}, Sec.2.2, paragraph 1, last sentence, quoted: "The histogram of values of $cos\phi$ ($\phi$ in 2D) weighted by the magnitude of the gradient at each voxel (pixel) is what we call HRO." }. The relative orientation is (i) $\zeta >0 $ corresponds to an HRO showing $B_{POS}$ predominantly perpendicular to $\nabla I$. (ii) $\zeta \sim 0 $ corresponds to a flat HRO showing no predominant relative orientation between $B_{POS}$ and $\nabla I$. (iii) $\zeta < 0 $ corresponds to an HRO showing $B_{POS}$ predominantly parallel to $\nabla I$.

\begin{figure}[t]
	\centering
	{\includegraphics[width=.99\linewidth,height=2.38\linewidth]{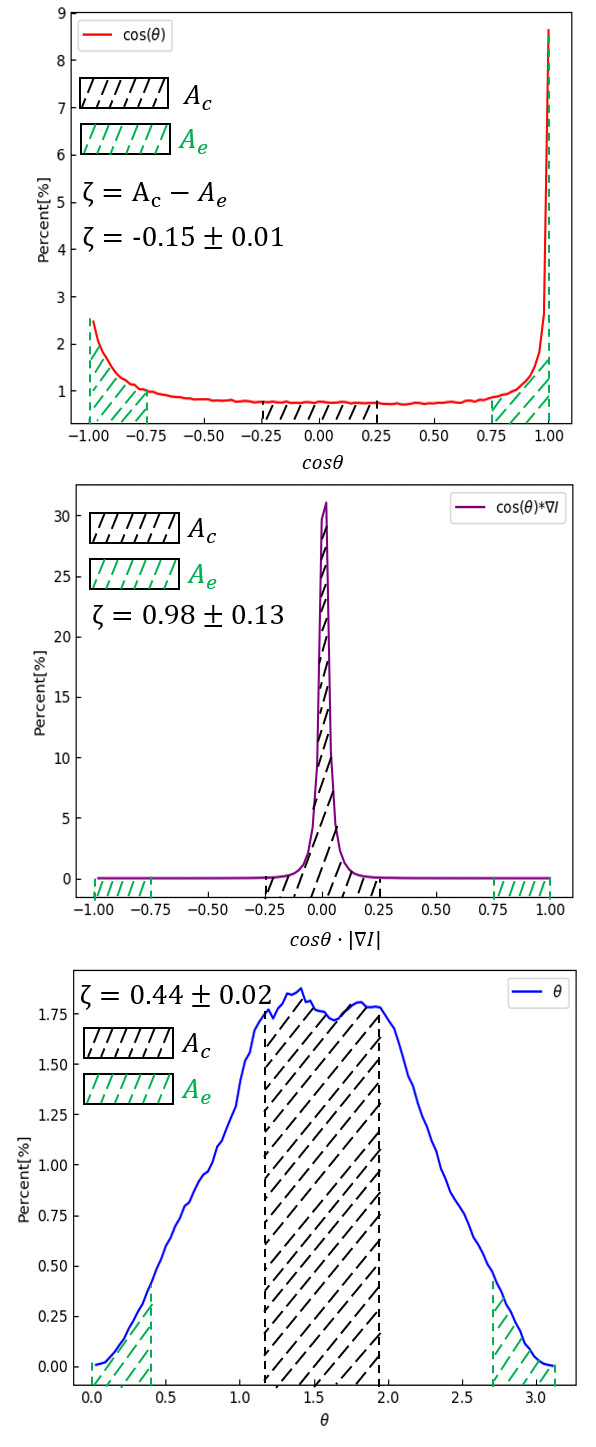}}
	\caption{\label{fig:cos} The the histograms of $cos (\theta$) (top), $cos (\theta)\cdot|\nabla I|$ (middle), and $\theta$ (bottom) plotted as HRO format, using simulation M$_A0.2$. $\zeta=A_c-A_e$ is the histogram shape parameter: , where and $A_c$ is the area under the central region of the HRO curve, $A_e$ is the area in the extremes of the HRO. }
\end{figure}

However, $cos (\theta)\cdot|\nabla I|$ does not appropriately reveal the information of spatially relative orientation in a histogram format. In Fig.~\ref{fig:cos}, we also plot the histograms of $cos (\theta$), $cos (\theta)\cdot|\nabla I|$, and $\theta$ using simulation M$_A0.2$. We see that the histogram of cos$\theta$ is not a Gaussian, but the histogram of $cos (\theta)\cdot|\nabla I|$ is shaped to be a Gaussian profile since the distribution of $cos(\theta)$ is dominated by the distribution of $|\nabla I|$, which is already a Gaussian itself \citep{YL17a,2018arXiv180200024Y}. The distribution of $|\nabla I|$ does not reveal the orientation of density structures.  We clearly see that the histogram of $cos(\theta)$ is not in Gaussian shape, with $\zeta=-0.15\pm0.01$. However, after weighted by normalized $|\nabla I|$, the histogram becomes Gaussian, with $\zeta=0.98\pm0.13$. Therefore, in terms of studying the relative orientation of $B_{POS}$ and $\nabla I$, the weighted histogram can give different results through the utilization of $\zeta$. 

We thus propose to remove the weighting term $|\nabla I|$ and using the histogram\footnote{For the histogram of $\theta$, $A_c$ is the area under the central region $\frac{3}{8}\pi<\theta<\frac{5}{8}\pi$, $A_e$ is the area in extreme region $0<\theta<\frac{1}{8}\pi$ and $\frac{7}{8}\pi<\theta<\pi$.} of $\theta$ instead of $cos\theta$. Since the transformation between $\theta$ and $cos(\theta)$ is not linear, the Gaussian profile of $\theta$ is deformed after being transformed into $cos(\theta)$. In this case, the utility of $\zeta$ might cause confusion, as $\zeta$ is highly sensitive to the shape of histograms. As shown in Fig.~\ref{fig:cos}, $\zeta$ illustrates different pictures for the relative orientation of $B_{POS}$ and $\nabla I$. The scientific consideration in Sec.~\ref{sec:theory} shows that globally $\nabla I$ tends to be perpendicular $B_{POS}$. The histogram of $\theta$ thus reveal more accurate physical structures for the relative orientation of $B_{POS}$ and $\nabla I$. An alternative way to accurately quantify the global relative orientation is the utility of AM, which is implemented in VGT and not sensitive to the shape of histograms (see Sec.~\ref{sec.method} for the definition of AM).

\begin{figure*}[t]
	\centering
	\includegraphics[width=.99\linewidth,height=.80\linewidth]{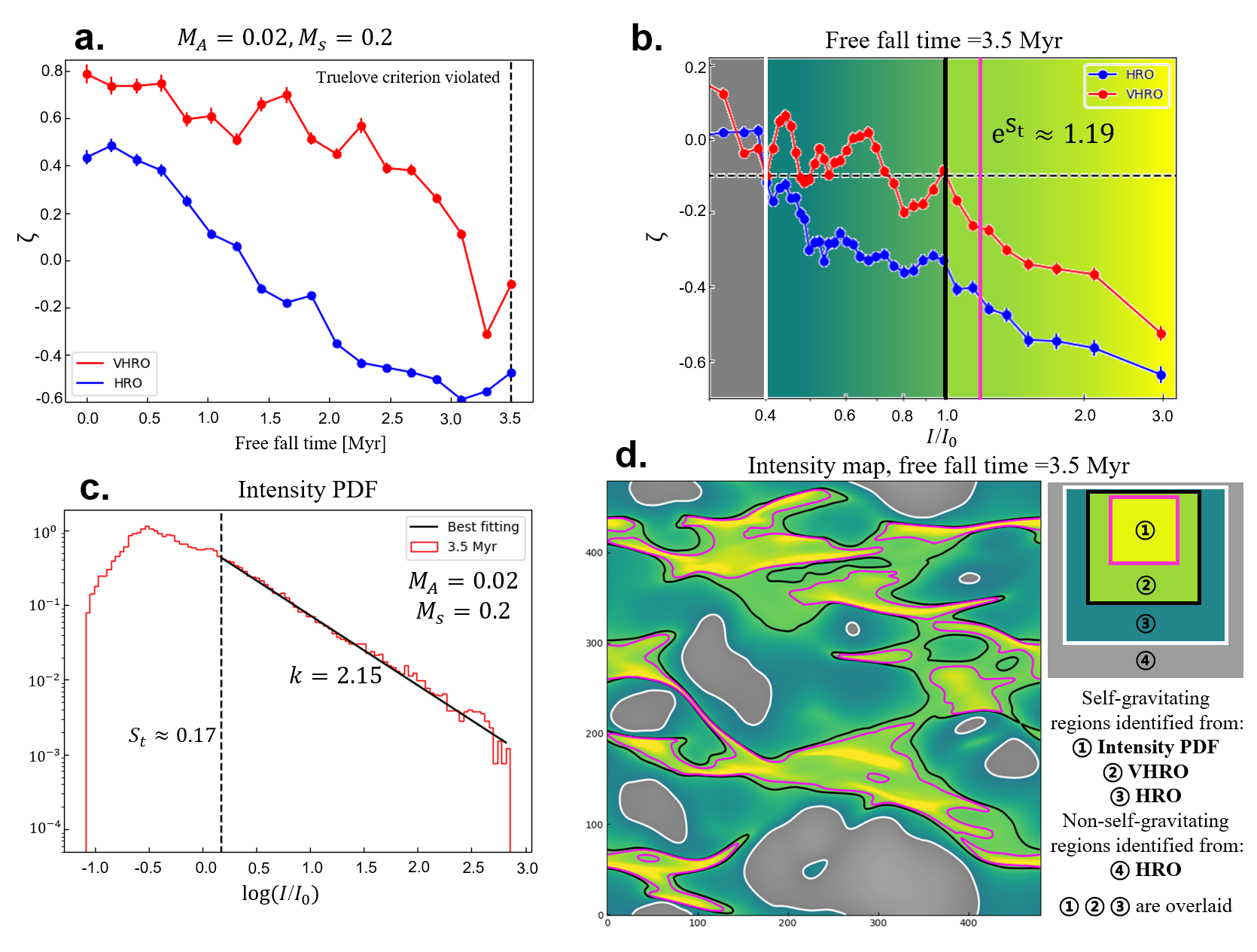}
	\caption{\label{fig:VHRO}\textbf{Panel a:} the variation of $\zeta$ in time after the self-gravity is turned on. The dashed line indicates the time when the Truelove criterion is violated. \textbf{Panel b:} the variation of $\zeta$ to different intensity segments $I/I_0$, where $I_0$ is the mean intensity. The dashed line indicates $\zeta=-0.1$. Green color represents the start point of HRO, while lime color represents the start point of VHRO. \textbf{Panel c:} The lognormal PDF plus power-law models. The dashed black line outlines all the density past the transition density S$_t$, which is self-gravitating. $k$ is the slope of power-law relation. \textbf{Panel d:} Intensity maps of the projection of simulation M$_s0.2$, free fall time 3.5 Myr. Overlaid colors correspond to regions with intensity shown in panel b.}
\end{figure*}
\subsection{Velocity Histogram of Relative Orientation}
In Sec.~\ref{subsec:HRO}, we modified HRO by removing the weighting term $|\nabla I|$ and using histogram of $\theta$. Based on these modifications, we make a synergy of VGT and HRO techniques, called Velocity Histogram of Relative Orientation (VHRO). The algorithm of VHRO is following HRO, but using velocity gradients instead of intensity gradients. The $\zeta$ is still implemented in VHRO for simplicity.

In Sec.~\ref{sec.obs}, we shows that intensity gradients may change their orientation to be parallel to magnetic fields at high-density regions in case of super-sonic turbulence. However,in the presence of self-gravity, \citet{YL17b} \& \citet{survey} pointed out that the matter infall induces a change of the direction of intensity/velocity gradients with respect to magnetic field. In other words, towards regions where star formation is taking place, the intensity/velocity induced by the infall motions parallel to the magnetic field gradually begin to dominate over the intensity/velocity arising from turbulence. As a result, both velocity gradients and intensity gradients are changing their direction by 90$^\circ$, thus becoming parallel to the magnetic field direction.
	
Fig.~\ref{fig:VHRO}a gives one numerical example of global intensity gradients analyzed by the modified HRO and global velocity gradients in thin channels analyzed by VHRO, in response to the increment of self-gravity. We choose to use the sub-sonic simulation M$_s$0.2, which has no contribution from shocks. We see that the $\zeta$ for both HRO and VHRO is decreasing as the free all time goes increasing. It means that the rotated intensity gradients and velocity gradients become perpendicular to magnetic fields with the increment of self-gravity. However, the change of $\zeta$ is more dramatically for intensity gradients. We expect the reason is that the change of density field is an accumulating process, while velocity field is significantly changed only when the gravitational energy dominates over the kinematic energy of turbulence. This thus provides a way of, first of all, locating regions dominated by self-gravity, and second, identifying the stage of gravitational collapsing for molecular clouds using the different behaviors of intensity gradients and velocity gradients\footnote{The change of relative orientation of velocity gradient and intensity gradient is also expected to happen in front of shocks. However, shocks can be distinguished from the self-gravity regions through morphological differences. For example, the curvature of the gradient field for the gravitational collapse is expected to be larger than the curvature of the gradient field for shocks.}.

In Fig.~\ref{fig:VHRO}b, we separate the intensity map \textbf{I(x,y)} at free all time 3.5 Myr into 40 segments. The intensity of the $n^{th}$ segment locates at the interval between the $2.5\cdot(n-1)$ and the $2.5\cdot n$ percentile of \textbf{I(x,y)}. We analysis the relative orientation of intensity/velocity gradients and magnetic fields through HRO/VHRO in each segment. $\zeta=0.0$ indicates the relative orientation tends to be neither parallel nor perpendicular, so we claim that when $\zeta<=-0.1$ the gradients and magnetic fields start changing their relative orientation. In Fig.~\ref{fig:VHRO}b, we find that the $\zeta$ of HRO is negatively proportional to the intensity in the corresponding segment. It indicates the intensity gradients are continuously changing their relative orientation from perpendicular to parallel to the magnetic fields. The critical intensity values, above which the change of relative orientation happens, is $I/I_0\approx 0.4$. However, as for the $\zeta$ of VHRO, we see there exists a transitional range between $I/I_0\approx 0.4$ and $I/I_0\approx 1.0$, at which the value of $\zeta$ is oscillating around $-0.1$. In the case of $I/I_0\ge 1.0$, the $\zeta$ of VHRO is monotonically decreasing. Therefore, it confirms that velocity gradients change their relative orientation to magnetic fields, only when gravitational energy starts dominating the turbulence system. 

In Fig.~\ref{fig:VHRO}c, we plot the gas intensity probability distribution function (PDF), which evolves to a combination of lognormal ($P_{N}$) PDF  at low intensities and a power-law ($P_{L}$) PDF  at high intensities in the case of self-gravitating MHD turbulence \citep{2019ApJ...879..129B}: 
\begin{equation}
    P_{N}(s)=\frac{1}{\sqrt{2\pi\sigma_s^2}}e^{-\frac{(s-s_0)^2}{2\sigma_s^2}},s\le S_t
\end{equation}
\begin{equation}
    P_{L}(s)\propto e^{-k s},s>S_t
\end{equation}
where s$=log(I/I_0)$ is the logarithmic intensity and $\sigma_s$ is the standard deviation of the lognormal, while $I_0$ and $s_0$ denote mean intensity and mean logarithmic intensity. S$_t$ is the logarithm of the normalized transitional intensity between lognormal and power-law forms of the intensity PDF. Fig.~\ref{fig:VHRO}c shows that when $s\ge S_t\approx0.17$, the PDF is following a power-low correlation with slope $k=2.15$. As a result, when $I/I_0$ gets larger than $e^{S_t}\approx 1.19$, the gas is expected to be self-gravitating.

In Fig.~\ref{fig:VHRO}d, we show the intensity contours corresponding to three critical intensity values: (i) $I/I_0\ge 0.4$ (i.e., intensity gradients start changing the relative orientation), green area enclosed by white contours; (ii) $I/I_0\ge 1.0$ (i.e., velocity gradients start changing the relative orientation), lime area enclosed by black contours. (iii) $I/I_0\ge 1.19$ (i.e., the gas triggers self-gravity), pink contours. We see that all lime areas embed in green areas, while pink contours all locates within the lime area. The close correspondence of self-gravitating regions and the regions obtained with velocity gradients reveals that velocity gradients  analysed by VHRO are sensitive in identifying self-gravitating regions. However, this is not the case for density gradients analysed by HRO. According to HRO, only small pieces are not gravitational collapsing, while in reality only small fraction is collapsing. We therefore conclude that VHRO is more powerful in identifying gravitational collapsing regions than HRO.

\section{Discussion}
\label{sec.dis}
\subsection{Density and velocity statistics} 
In MHD turbulence, density and velocity fluctuations show different statistical behaviors. The velocity fluctuations follow the same GS95 relation for Alfvenic turbulence, while for density fluctuations the GS95 relation could be valid only by filtering out high contrast density clumps \citep{2005ApJ...624L..93B}. The studies of density and velocity fields, therefore, provide different information about MHD turbulence and magnetic fields. For example, \citet{2007ApJ...658..423K, YL17b,2019arXiv190506341X} show that the high contrast density fluctuations are compressed by shocks perpendicular to the local direction of the magnetic field, while velocity structures always remain aligned with their local magnetic fields. As a result, the density gradients and velocity gradients become perpendicular in front of shocks. Without relying on polarimetry, the study of density gradients and velocity gradients thus provides a possible method for identifying shock structures.

The LP00 theory shows that it is possible to change the relative contribution of density and velocity in PPV cubes. LP00 suggested that density fluctuation dominates in thick velocity channels, while velocity fluctuation dominates in thin velocity channels. This assumption holds not only in the single-phase self-absorption media, but also in two-phase \ion{H}{1} media \citep{LP04,LP06,LP08,2017MNRAS.464.3617K,2010ApJ...714.1398C}. Therefore, by varying the thickness of velocity channels, one can both trace the magnetic field and identify shocks in diffuse regions and molecular clouds. 

\subsection{Contribution from thermal broadening effect}
We propose IGT as a tool complementary to the VGT technique. The latter technique has proven successful both for studies magnetic fields
in atomic hydrogen \citep{YL17a,2018ApJ...865...59L,2018ApJ...865...46L,2019ApJ...874...25G,2018MNRAS.480.1333H}, and molecular clouds \citep{survey, Laura}. While gas in molecular clouds presents one phase media, the existence of two phases of \ion{H}{1} induced some researchers to question the validity of the interpretation of the results obtained with the VGT in terms of velocity gradients. In particular, \cite{2019arXiv190201409C} claim that the structures observed in thin velocity channel maps arise from actual density structures rather than the velocity caustics as they were interpreted in the papers as mentioned earlier \citep{LP04, LP06, LP08,2017MNRAS.464.3617K,2010ApJ...714.1398C}. If this were true, there would be no differences between the studies of IGs within thick slices and velocity gradients using thin slices, i.e., using VChGs technique \citep{2018ApJ...865...59L}. The results in the present paper contradict this conclusion. First of all, the maps of gradients and the AM obtained with the VChG technique are very different from those obtained with IGs, but similar to those obtained with velocity centroid gradients.
Moreover, the regions where the directions obtained with IGs and VChGs are different coincide with the shock regions (see Fig.~\ref{fig:serpens}) and the directions obtained with VChGs and IGs in these regions are close to 90$^\circ$. The latter are the expectations of the gradient theory based on the modern understanding of MHD turbulence \citep{BL19} and, at the same time, these facts are difficult to explain assuming the structures in the thin and thick velocity channel maps arise from actual enhancements of underlying hydrogen density. Our reply to \cite{2019arXiv190201409C} is made public in \citet{2019arXiv190403173Y} and below we explain why we believe that the measurements of intensities in thin and thick channels deliver velocity and density information respective.

\cite{2019arXiv190201409C} uses both GALFA-\ion{H}{1} observational data and numerical simulations to address the physical nature of thin velocity channels in PPV space. The study questions the validity and applicability of the statistical theory of PPV space fluctuations formulated in \citet{LP00}, denoted as LP00 later, to \ion{H}{1} gas. They concludes that (i) the thermal broadening effect washes out the velocity information in thin velocity channels, in the case of sub-sonic turbulence; (ii) the structure in thin velocity channels arises from density fluctuations rather than velocity fluctuations, in the case of super-sonic turbulence; (iii) the observed change of spectral index with the evolution of the slice thickness is a consequence of two-phase medium effects.

The arguments about the thermal broadening effect raised in \cite{2019arXiv190201409C} are based on two-phase \ion{H}{1} medium. However, LP00 have already explicitly accounted for thermal broadening and evaluated its effect for both subsonic and supersonic turbulence. They found the thermal broadening effect gives a little contribution to the velocity information in thin velocity channels. Also, the observed change of spectral index is reported by different groups to be the same both in two-phase \ion{H}{1} medium and one phase media of CO isotopes \citep{1993MNRAS.262..327G, LP06,2000ApJ...543..227D,LP04,2001ApJ...551L..53S,2001ApJ...561..264D,2006ApJS..165..512K,2006AIPC..874..301L,2006MNRAS.372L..33B,2006ApJ...653L.125P}. Later, \citet{2019arXiv190403173Y} argued that the spectral indexes of velocity spectrum obtained with LP00 correspond to the expectation of MHD turbulence theory in both observation and numerical simulation. They also illustrate that the computation of correlation between PPV slices and dust emission in \cite{2019arXiv190201409C} is not sensitive in revealing the relative significance of velocity and density fluctuations in velocity channel maps. 

It was shown that strong shocks provide density structures are perpendicular to magnetic fields, while low-density filament structures formed by the shearing fluid are parallel to magnetic fields \citep{2019arXiv190506341X,YL17b,2005ApJ...624L..93B}. In the case of low M$_s<1$ turbulence, therefore, density structures without the presence of shocks are parallel to magnetic fields, in particular for high Galactic latitude regions. As a result, it is not surprising that \cite{2019arXiv190201409C} got structural similarity between Planck 857 GHz dust emission map and \ion{H}{1} thin channel maps at high Galactic latitude regions ($b>60^\circ$), since both density structures and velocity structures are parallel to magnetic fields.
		
In any case, here we see that IGs also can trace magnetic fields, while VChGs shows higher accuracy. Therefore, even in the situations when there are significant contributions from density, the validity of VChGs as a technique to trace magnetic field is not affected. \\

\subsection{Extracting 3D magnetic field structures} 
Due to the position of the solar system within the Galactic disk, the line of sight inevitably crosses more than one molecular cloud. It is therefore impossible to use far-infrared polarimetry to study the local magnetic fields in most molecular clouds. Fortunately, VGT and IGT show advantages in dealing with multi-clouds issues. In general, VChGs shows higher accuracy than IGs in terms of magnetic fields tracing. One possible reason is that the high-density clumps do not show the Goldreich-Sridhar type's anisotropy, although the density structure is always anisotropic at small scales with the presence of strong magnetic fields \citep{2005ApJ...624L..93B}. We thus expect that we can improve the performance of IGs in tracing magnetic fields by removing contrast density clumps or low spatial frequencies. In addition, as M$_A$ increases, the magnetic field along the line of sight varies rapidly, especially when the turbulence becomes super-Alfvenic. In this case, it is also important to remove the low spatial frequencies component. 
		
\citet{2018arXiv180909806H} showed the availability of gradients in tracing magnetic fields using synthetic molecular line maps of CO isotopologue. After that, \citet{Laura} demonstrated the utility of VChGs technique by using observational data from multiple molecular tracers to construct a 3D magnetic field structure. With the improved IGs, we expect to be able to apply it to 3D magnetic field construction using multiple molecular tracer maps in similarly to VChGs. One such method would be to stack the intensity gradient maps from $^{12}$CO, $^{13}$CO, C$^{18}$O to create a three-layer tomography map.

The galactic rotation curve can be used to isolate different clouds, including both diffuse \ion{H}{1} and molecular clouds, in the velocity space and allowing magnetic fields to be studied separately \citep{2019ApJ...874...25G}. It, therefore, opens new prospects for studying the 3D magnetic field structures in the Milky Way using IGs and VChGs.

\section{Conclusion}
\label{sec.con}
Based on the theory of MHD turbulence and turbulent reconnection, we show that the Intensity Gradients Technique, i.e., gradients calculated within thick velocity channel maps, can reveal the magnetic field orientation and magnetization in diffuse media and identify shock structures. The essence of the technique is to vary the channel thickness to change relative contribution from density and velocity statistics in PPV space. The gradients of thick channel maps carry information about the turbulent intensity fluctuation, while the gradients of the thin channel maps contain information about the turbulent velocity fluctuation. We compare the abilities of IGs and earlier proposed technique VChGs, as well as make synergy with HRO. To summarize:
\begin{enumerate}
		\item  The varying thickness of velocity channels changes the relative contribution from density and velocity statistics. The VChGs calculated within thin velocity channels contains more information about velocity statistics, while the IGs calculated within thick velocity channels contains more information about density statistics.
		\item We show that
	\begin{enumerate}
			\item The dispersion of intensity and velocity gradient distributions are applicable to reveal the magnetization in diffuse media.
			\item VChGs and RVCGs are more accurate than IGs in terms of tracing the magnetic field orientation. We propose to trace magnetic field orientation using VChGs for super-sonic turbulence, while RVCGs for sub-sonic turbulence.
			\item IGs tend to be parallel to its local magnetic fields when getting close to the dense shock front in the absence of gravity. IGs, therefore, has the advantage of identifying shock structures, while there is no particular universal density at which the change of the relative orientations happens.
	\end{enumerate}
	\item We apply IGs and VChGs to the GALFA-\ion{H}{1} data and get statistically similar results. In terms of magnetic fields tracing, VChGs shows better alignment with the magnetic field inferred from the Planck 353GHz polarimetry data. 
	\item We claim that IGT can be used synergetically with VGT for magnetic fields studies when velocity information is not available, for example, the \ion{H}{1} column density data.
	\item We demonstrated the advantages of the synergistic utility of different types of gradients (e.g., IGs, VChGs, RVCGs). We show the possibility of studying magnetic field ecosystems, shocks, and self-gravitational collapse, Alfvenic Mach number.
	\item we demonstrate significant differences between HRO and IGT. In particular, IGT is a technique to be used in conjunction with VGT, without employing polarimetry, while  HRO critically depends on polarimetry. 
    \item Our work shows how to utilize intensity gradients using the procedures we developed earlier for velocity gradients. We also show that:
    \begin{enumerate}
        \item  velocity gradients can be used in a way similar to the modified HRO technique for density gradients. The proposed velocity-HRO (VHRO) can be used for identifying the regions of self-gravitating.
        \item velocity gradients starts change their relative orientation to magnetic fields, when gravitational energy starts dominating the turbulence system. 
        \item Self-gravitating regions embed in the transition regions obtained with velocity gradients. Velocity gradients are, therefore, sensitive in identifying gravitational collapse.
    \end{enumerate}
\end{enumerate}
	
{\bf Acknowledgment}  {A.L. acknowledges the support of the NSF grant AST 1715754, and 1816234. K.H.Y. acknowledges the support of the NSF grant AST 1816234. Y.H. acknowledges the support of the NASA TCAN 144AAG1967. This publication utilizes data from Galactic ALFA \ion{H}{1} (GALFA-\ion{H}{1}) survey data set obtained with the Arecibo L-band Feed Array (ALFA) on the Arecibo 305m telescope. }
{\software ZEUS-MP/3D code \citep{2006ApJS..165..188H}}
	

	
	

	
\end{document}